\documentclass{WileyMSP-template}
\usepackage{color}

\usepackage{ragged2e}

\begin{document}

\pagestyle{fancy}
\rhead{\includegraphics[width=2.5cm]{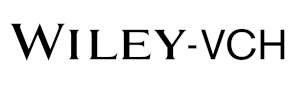}}

\title{Irrelevant corrections at the quantum Hall transition}

\maketitle

\author{Keith Slevin}
\author{Tomi Ohtsuki}

\begin{affiliations}
Keith Slevin\\
Department of Physics, Graduate School of Science, Osaka University, Toyonaka, Osaka 560-0043, Japan\\
Email Address:slevin.keith.sci@osaka-u.ac.jp

Tomi Ohtsuki\\
Division of Physics, Sophia University, Chiyoda, Tokyo 102-8554, Japan\\
Email Address:ohtsuki@sophia.ac.jp

\end{affiliations}


\keywords{quantum Hall effect, critical exponent, finite size scaling}

\begin{abstract}
The quantum Hall effect is one of the most extensively studied topological
effects in solid state physics.
The transitions between different quantum Hall states exhibit critical phenomena
described by universal critical exponents.
Numerous numerical finite size scaling studies have focused on the critical exponent $\nu$ for the 
correlation length. 
In such studies it is important to take proper account of irrelevant corrections to scaling.
Recently, Dresselhaus et al. [Phys. Rev. Lett. {\bf 129}, 026801(2022)] proposed a
new scaling ansatz, which they applied to the two terminal conductance of the Chalker-Coddington model.
In this paper their proposal is applied to our previously reported data for the Lyapunov exponents of that model
using both polynomial fitting and Gaussian process fitting.
\end{abstract}

\justifying

\section{Introduction}
Four decades after its discovery\cite{Klitzing80}, the quantum Hall effect continues to attract considerable
attention, both experimentally and theoretically.
Its most striking feature is the plateau behaviour of the Hall conductance $\sigma_{yx}$,
which is quantised in integer multiple of $e^2/h$
\begin{equation}
    \sigma_{yx}=k \frac{e^2}{h}\,.
\end{equation}
Here $h$ is Planck's constant,  $e$ is the elementary charge, and $k$ is an integer topological number \cite{Thouless82}.
As a control parameter such as two dimensional electron density (i.e., gate voltage) or perpendicular magnetic field
is varied, the Hall conductance $\sigma_{yx}$ changes from one quantised value to the other.
The transition is accompanied by a power law divergence of a correlation length $\xi$,
\begin{equation}
    \xi \sim \frac{A}{|x-x_c|^\nu}\,,
\end{equation}
with $A$ a constant and a critical exponent $\nu$ that is expected to have a universal value.

In early finite size scaling analyses, the value of $\nu$ was estimated to be $\approx 2.3$ \cite{Huckestein95,Evers2008RMP}.
The problem was later revisited and a significantly higher estimate $\nu\approx 2.59$ \cite{Slevin09,Amado11,Fulga11,Dahlhaus11,Obuse12,Slevin12,Gruzberg17,Puschmann19} was obtained  
(although a slightly smaller value, $\nu\approx 2.5$, was reported for an analysis of the Chern number \cite{Zhu19}).
The recent proposal of a conformal field theory \cite{Bondesan17,Zirnbauer19} with 
 marginal scaling, i.e., with $\nu=\infty$, has  stimulated further numerical study
of this problem\cite{Dresselhaus21,Dresselhaus22}.

A major difficulty in finite size scaling analyses of the quantum Hall transition is the presence of slow
irrelevant corrections that complicate the estimation of the critical exponent.
In Ref.~\cite{Dresselhaus22}, the energy $x$ and system size $L$ dependence of
two terminal conductance $g(x,L)$ of the Chalker-Coddington model\cite{Chalker88,Kramer2005rev} was calculated
and a finite size scaling analysis performed.
In that paper, Dresselhaus et al. proposed a method for taking into account the irrelevant correction based 
on a factorisation ansatz for the scaling function.
In this paper, we re-analyse our previously published data for the Lyapunov exponents of the 
Chalker-Coddington model following their proposal.

\section{Method}

The data we re-analyse here is for the smallest positive Lyapunov exponent $\gamma$ 
of the Chalker-Coddington model reported in Refs. \cite{Slevin09,Slevin12}.
The simulations used the transfer matrix method\cite{Pichard1981,MacKinnon81,MacKinnon83}.
A detailed explanation of this method is given in Ref.~\cite{Slevin2014}.

Very long quasi-one-dimensional strips were simulated.
The parameters in the simulation are the transverse width $L$ of the strip and the energy $x$ 
relative to the centre of the Landau level measured in units of the Landau bandwidth.
Here, $L$ is equal to the number of saddle points in the transverse direction of the strip.
It is known that for this model the critical point is at $x=0$.

Simulations were performed for $x\in \left[-0.225,0.225\right]$ and 
$L = 4, 6, 8, 12, 16, 24, 32, 48, 64, 96, 128, 192$ and $256$.
The output of each simulation was the estimate of $\gamma$ and the 
standard deviation of that estimate for the given parameters $x$ and $L$.
The precision of the data for the Lyapunov exponents is 0.03\% except for $L=192$ and $256$ where the precision is 0.05\%.
Further details can be found in Refs.~\cite{Slevin09,Slevin12}.

The analysis of the data involves fitting the system size $L$ and energy $x$ dependence of the 
dimensionless quantity $\Gamma = \gamma L$ to the finite size scaling form including corrections to
scaling due to an irrelevant scaling variable \cite{Huckestein95,Slevin1999correction}.
\begin{equation}
    \Gamma \left(x, L \right) = F\left(\phi_1\left(x\right) L^{\alpha_1},\phi_2 \left(x\right) L^{\alpha_2} \right)\,,
\end{equation}
where $\phi_1$ and $\phi_2$ are, respectively, the relevant and irrelevant scaling variables.
These are distinguished by the sign of the corresponding exponent $\alpha_1 > 0$ and $\alpha_2 < 0$.
The scaling variables $\phi_1$ and $\phi_2$ are analytic functions of the energy $x$.
In addition, the relevant variable is zero at the critical point.

The method proposed by Dresselhause et al. is based on two assumptions. 
First is that the scaling function $F$ factorises as follows
\begin{equation} \label{eq:F}
\Gamma \left(x, L \right) =   F_1\left(\phi_1\left(x\right) L^{\alpha_1}\right)  F_2\left(\phi_2\left(x\right) L^{\alpha_2}\right) \,.
\end{equation}
The second is that the $x$ dependence of the irrelevant variable can be neglected.
With these assumptions
\begin{equation}
\Gamma \left(x, L \right) =   F_1\left(\phi_1\left(x\right) L^{\alpha_1}\right)  F_2\left(\phi_2 L^{\alpha_2}\right) \,.
\end{equation}
Dresselhause et al. then note that the following ratio,
\begin{equation} \label{eq:GammaTilde}
\tilde{\Gamma}  \left(x, L \right) = \frac{ \Gamma \left(x, L \right) }{\Gamma \left(x=0, L \right)} =  \frac{ F_1\left(\phi_1\left(x\right) L^{\alpha_1}\right) }
{F_1 \left( 0 \right)}\,,
\end{equation}
obeys a single parameter scaling law, i.e., that the correction due to the irrelevant variable drops out of the analysis.

If we set $x=0$ in Eq. (\ref{eq:F}) we find
\begin{equation}
    \Gamma \left(x=0, L \right)   = F \left(0,\phi_2 \left(0\right) L^{\alpha_2} \right)\,.
\end{equation}
Thus, one way that the presence of a correction due to an irrelevant variable should be manifest is a system size dependence of
$\Gamma$ at the critical point.
Such a dependence is clearly visible in the data from our simulations cf. Fig 2 of Ref.~\cite{Slevin12}.
This system size dependence is removed from $\tilde{\Gamma}$ since 
\begin{equation} 
\tilde{\Gamma}  \left(x=0, L \right) =  1 \,,
\end{equation}
by definition.
However, this also means that the quantity  
\begin{equation}
    \Gamma_c = \lim_{L\rightarrow \infty} \Gamma \left(x = 0, L \right) \,,
\end{equation}
which is related to the multi-fractal properties of the critical wavefunctions
\cite{Chalker88,Cardy,Janssen,Evers2008RMP},
cannot be estimated by fitting $\tilde{\Gamma}$.

For brevity in what follows, when fitting data for $\tilde{\Gamma}$ we drop the suffix from the relevant variable and write simply
\begin{equation} \label{eq:ScalingLaw}
\tilde{\Gamma}  \left(x, L \right) = F \left(\phi\left(x\right) L^{\alpha}\right) 
\,,
\end{equation}
where the constant $F_1\left(0\right)$ has been absorbed into a re-definition of the scaling function $F$.
The correlation length critical exponent $\nu$ is related to $\alpha$ by
\begin{equation}
    \nu = \frac{1}{\alpha}\,.
\end{equation}

\section{Numerical results}

In this section we report the results of finite size scaling fits 
to the data for $\tilde\Gamma$ obtained as described above.
The standard deviations for the estimates of $\tilde\Gamma$ are calculated as described in the 
Appendix.
The fractional error is in the range 0.04\% to 0.07\%.
Since, for $x=0$, $\tilde\Gamma=1$ by definition, data points for this energy are not used in the fitting.
In our simulation, periodic boundary conditions were imposed in the transverse direction.
The Lyapunov exponent is then an even function of $x$.
We report results for both polynomial fitting and Gaussian process fitting.

\subsection{Polynomial fitting}

To ensure that the fitting function is an even function of $x$ we make $F$ an even function and $\phi$ an odd function.
Both $F$ and $\phi$ are expanded in Taylor series.
The series are truncated at orders $n$ and $m$ respectively.
Thus, we arrive at
\begin{equation}
\label{eq:TaylorExpansion}
    F\left(\phi L^\alpha \right) = 1 + \left( \phi L^\alpha \right)^2 + a_4 \left( \phi L^\alpha \right)^4 + \cdots + a_n \left( \phi L^\alpha \right)^n\,,
\end{equation}
and
\begin{equation}
    \phi \left(x\right) = b_1 x + b_3 x^3 + \cdots b_m x^m \,.
\end{equation}
The fitting parameters are the coefficients in the truncated Taylor series and the critical exponent $\nu$.
We fix $a_2=1$ to avoid ambiguity in the parameterisation and ensure universality of $F$.
The total number of parameters is
\begin{equation}
    N_P = 1+\frac{m+1+n-2}{2} \,.
\end{equation}
The best fit is found by minimising the $\chi$-squared statistic.
The quality of the fit is evaluated using the goodness-of-fit probability $p$.
Fits with $p<0.05$ are rejected.
Both $p$ and the standard deviations for the parameters were determined by generating and fitting 500 synthetic data sets.
Details of this method are given in Ref.~\cite{Slevin2014}.

The results of polynomial fitting are tabulated in Table 1. We found that it was not possible to fit all 
our simulation data using the ansatz of Dresselhaus et al. 
When data for systems sizes $L\le 32$ were excluded acceptable fits were obtained, i.e., goodness-of-fit $p\ge 0.05$.
When data with energy $x > 0.175$ were excluded it was possible to fit our data when data for $L=32$ were also included.

In Figure 1 we plot the data and the finite size scaling fit corresponding to the first line of Table 1.
Note that the error in the data is much smaller than the symbol size.
In Figure 2 we plot the same data as a function of the relevant scaling variable $\phi$ to demonstrate the 
collapse of the data onto a single curve that characterises single parameter scaling.

In Figures 3 and 4 we do the same for the fit corresponding to the last line of Table 1 where data for $L=32$ are also included
but the energy range is restricted.
The estimate of the exponent varies by about 1\% or so depending on the data included and the
order at which the Taylor expansions of $F$ and $\phi$ are truncated.
It is noticeable that excluding non-linearity of the relevant scaling variable leads to slightly lower estimates of the critical exponent.

\begin{table}
 \caption{Results of polynomial fitting: the estimate of the critical exponent $\nu$ and the standard deviation of this estimate together with details of the fit including the number of data points $N_D$, the number of parameters $N_P$, the minimum value of $\chi$-squared, 
 the goodness-of-fit probability $p$, and the orders $m$ and $n$ of the Taylor expansions.}
  \begin{tabular}[htbp]{@{}cccccccccc@{}}
    \hline
     &  & $\nu$ & $N_D$ & $N_P$ & $\chi^2$ & $p$ & $m$ & $n$ \\
    \hline
    $L\ge 48$  & $\left|x\right|\le 0.225 $  & $2.560\pm.002$ & 156 & 5 & 165 & 0.2 & 3 & 6  \\
    $L\ge 48$  & $\left|x\right|\le 0.225 $  & $2.568\pm.004$ & 156 & 7 & 157 & 0.3 & 5 & 8  \\
    $L\ge 48$  & $\left|x\right|\le 0.175 $  & $2.545\pm.001$ & 116 & 3 & 123 & 0.3 & 1 & 4  \\
    $L\ge 48$  & $\left|x\right|\le 0.175 $  & $2.562\pm.004$ & 116 & 5 &  98 & 0.8 & 3 & 6  \\
    $L\ge 32$  & $\left|x\right|\le 0.175 $  & $2.542\pm.001$ & 138 & 5 & 160 & 0.1 & 1 & 8  \\
    $L\ge 32$  & $\left|x\right|\le 0.175 $  & $2.552\pm.003$ & 138 & 5 & 150 & 0.2 & 3 & 6  \\
    \hline
  \end{tabular}
\end{table}

\begin{figure}
  \begin{center}
  \includegraphics[width=0.75\linewidth]{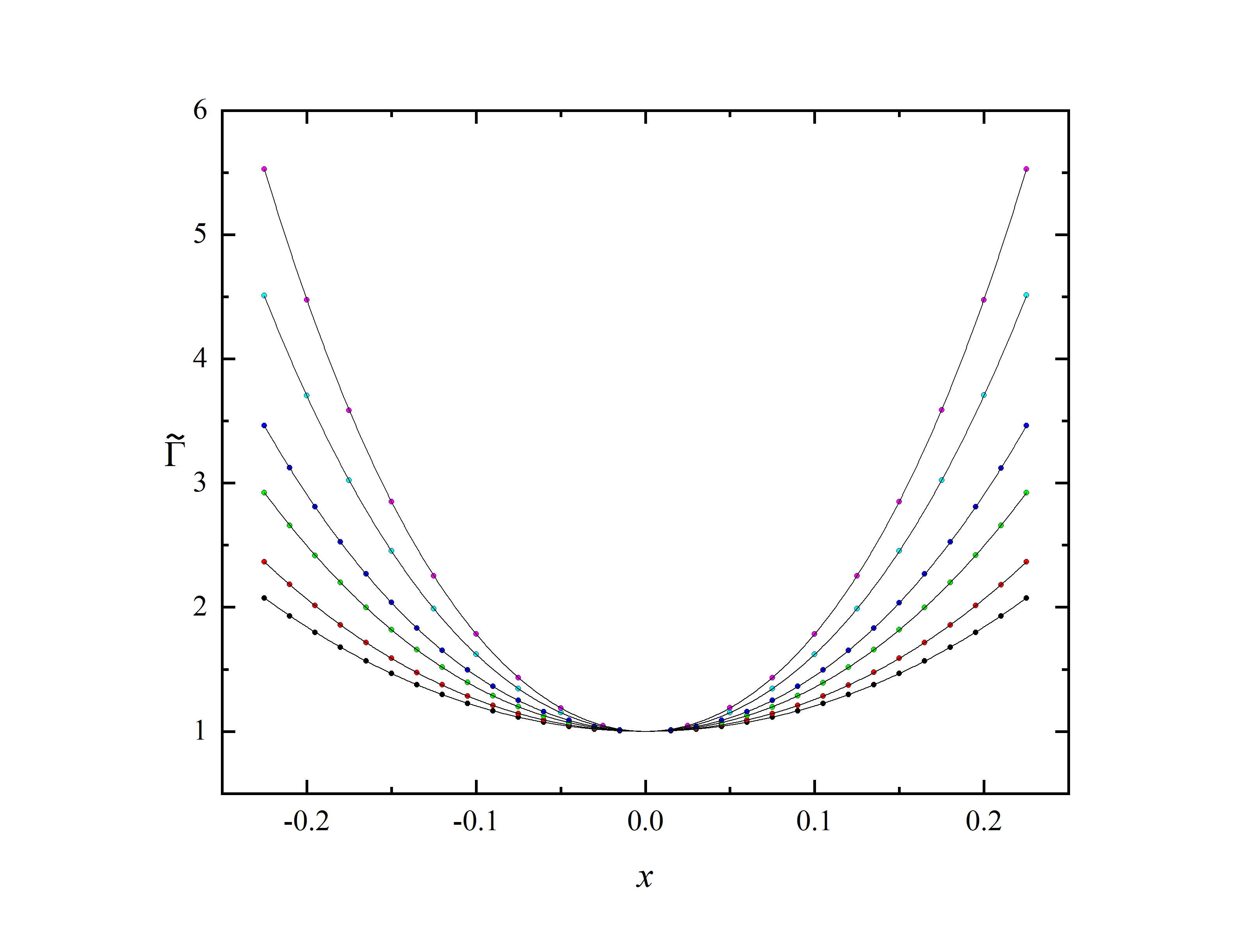}
  \end{center}
  \caption{Polynomial fit (lines) to data (points) for $L\ge48$. The details of the fit are given in the first line of Table 1.}
  \label{fig:1}
\end{figure}

\begin{figure}
  \begin{center}
  \includegraphics[width=0.75\linewidth]{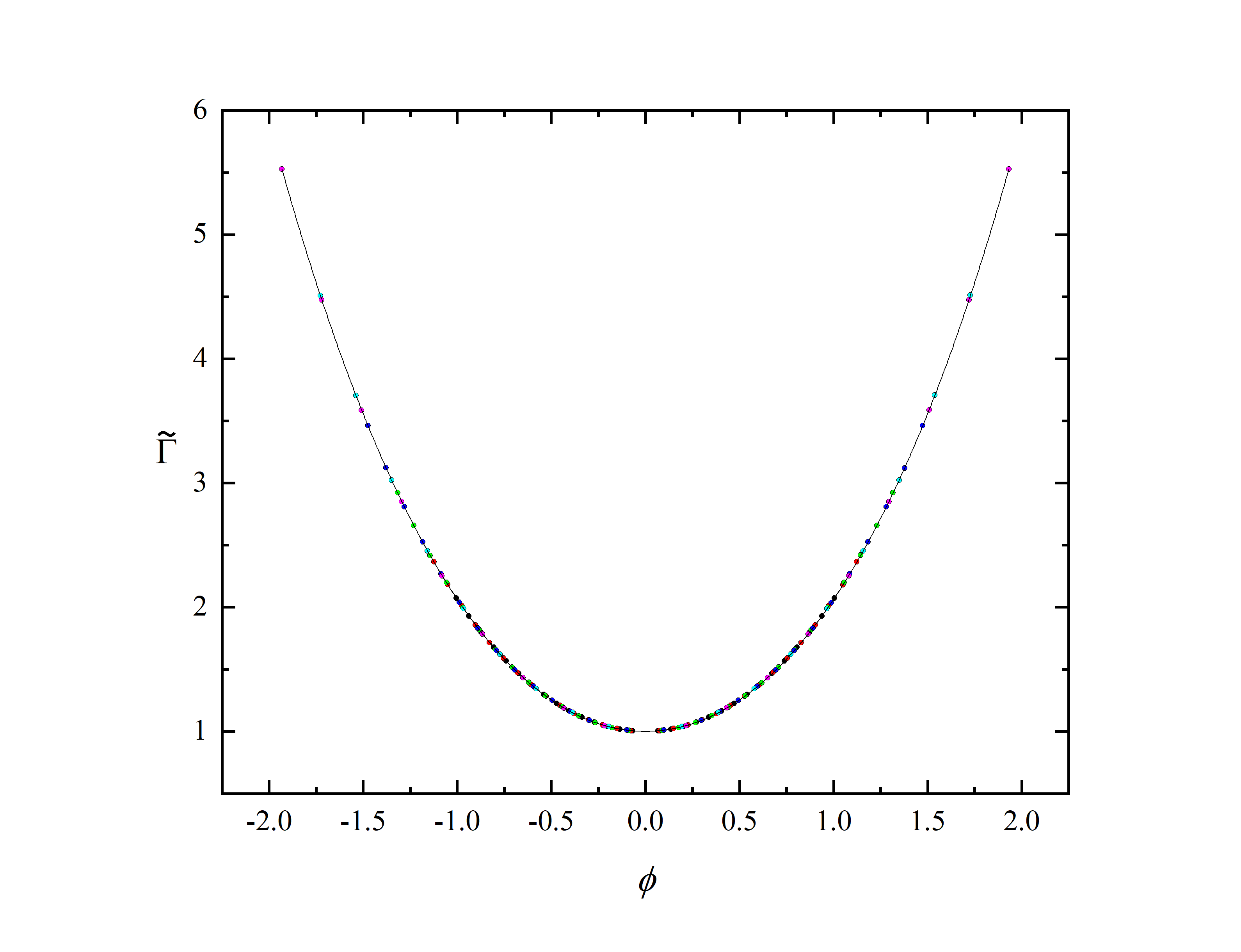}
   \end{center}
  \caption{Data (points) and scaling function (line) to demonstrate single parameter scaling of the data shown in Figure 1.}
  \label{fig:2}
\end{figure}

\begin{figure}
  \begin{center}
  \includegraphics[width=0.75\linewidth]{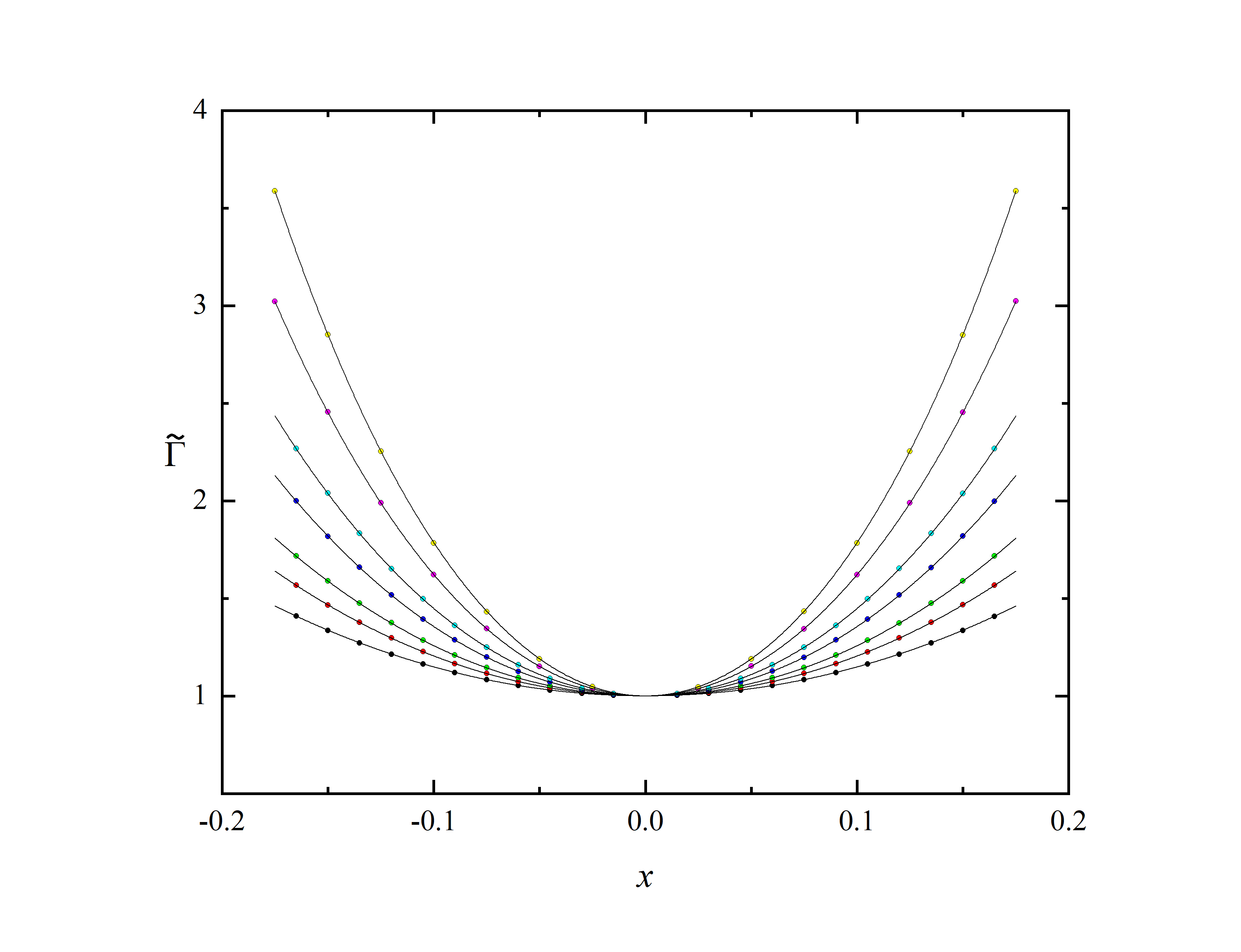}
  \end{center}
  \caption{Polynomial fit (lines) to data (points) for $L\ge32$. The details of the fit are given in the last line of Table 1.}
  \label{fig:3}
\end{figure}

\begin{figure}
  \begin{center}
  \includegraphics[width=0.75\linewidth]{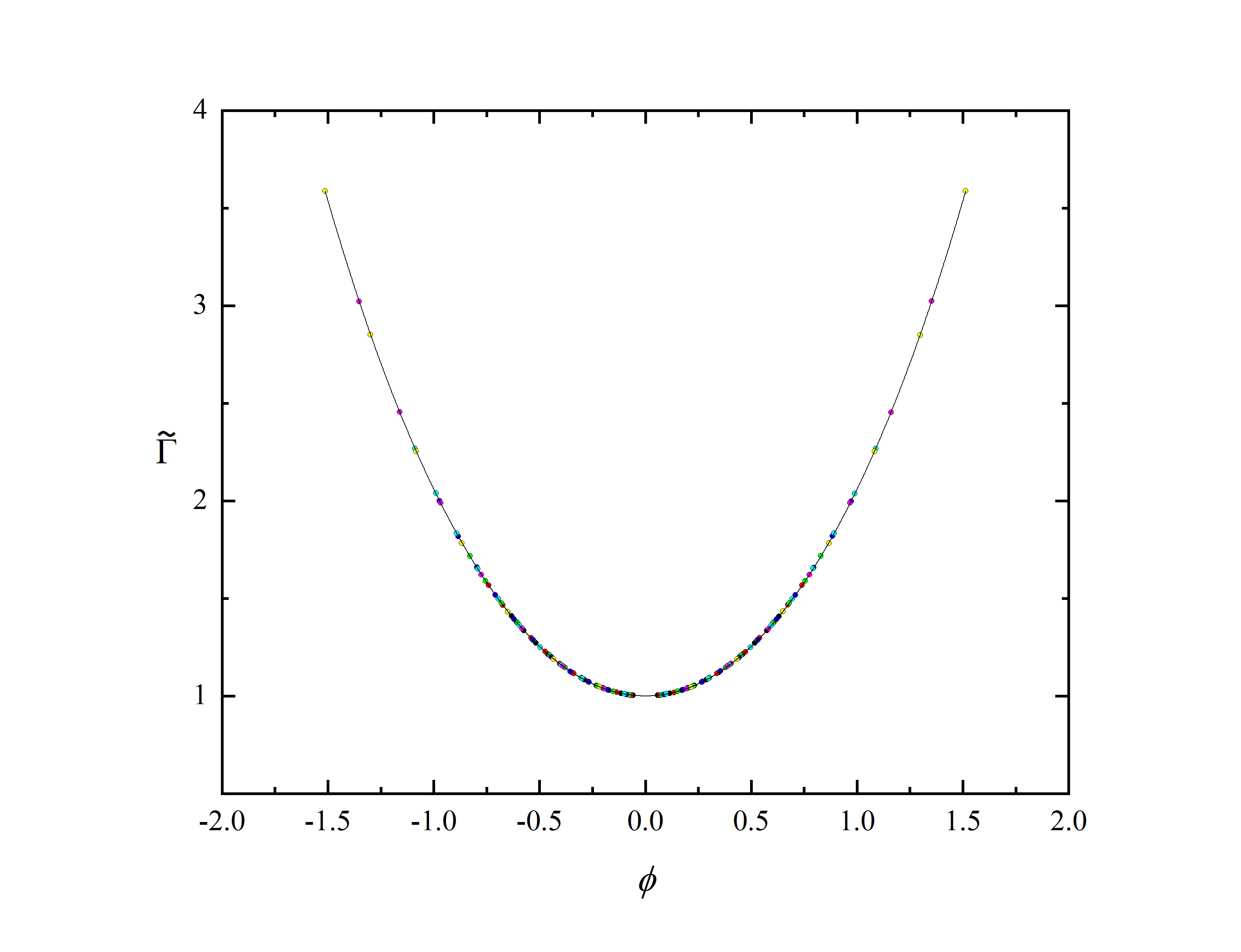}
   \end{center}
  \caption{Data (points) and scaling function (line) to demonstrate single parameter scaling of the data shown in Figure 3.}
  \label{fig:4}
\end{figure}

\subsection{Gaussian process fitting}
For comparison we also report the results of Gaussian process fitting \cite{Harada11} of the data for $\tilde{\Gamma}$.
In this method a Gaussian process is used to represent the scaling function $F$ rather than a polynomial.
Also non-linearity of the relevant scaling variable $\phi$ with respect to $x$ is neglected, i.e., $\phi=x$.
In this fit, in contrast to the polynomial fitting [Eq.~(\ref{eq:TaylorExpansion})],
the condition $F(0)=1$ is not imposed.
There are 4 fitting parameters: the critical exponent $\nu$, and three parameters that characterise the Gaussian process.
These parameters are varied to maximise the likelihood function \cite{Harada11,Yoneda20}.
The standard deviation for the estimate of the exponent was obtained from 1000 Monte Carlo samples.
Unfortunately a goodness-of-probability $p$ is not available for this method of fitting.

The results are tabulated in Table 2 and the scaling collapse of the data is shown in Figure 5.
When the range of energy $x$ is restricted so that a polynomial fit that neglects non-linearity of the scaling variable 
is acceptable, i.e., $p>0.05$ for the polynomial fit, the estimates of $\nu$ for both methods agree.
However, when the range of energy $x$ is widened so that non-linearity must be included to obtain an acceptable goodness-of-fit for the polynomial fit, the polynomial fit gives a larger estimate of $\nu$ than the Gaussian process fit.
Comparing the first lines of Table 1 and Table 2 we see that non-linearity, which is included in the polynomial fit but not 
in the Gaussian process fit, gives a noticeably lower $\chi$-squared.

\begin{table}
 \caption{Results of Gaussian process fitting: the estimate of the critical exponent $\nu$ and the standard deviation of this estimate together with details of the fit including the number of data points $N_D$, and the minimum value of $\chi$-squared.}
\label{tab:gaussian}
  \begin{tabular}[htbp]{@{}cccccc@{}}
    \hline
      & &$\nu$   & $N_D$& $\chi^2$  \\
    \hline
$L\ge 48$& $|x|\le 0.225$ & $2.533\pm 0.001$ & 156 & 353  \\
$L\ge 48$& $|x|\le 0.175$ & $2.545\pm 0.001$ & 116 & 117 \\
$L\ge 32$& $|x|\le 0.175$ & $2.542\pm 0.001$ & 138 & 160 \\
    \hline
  \end{tabular}
\end{table}

\begin{figure}
  \begin{center}
  \label{fig:gaussian}
  \includegraphics[width=0.75\linewidth]{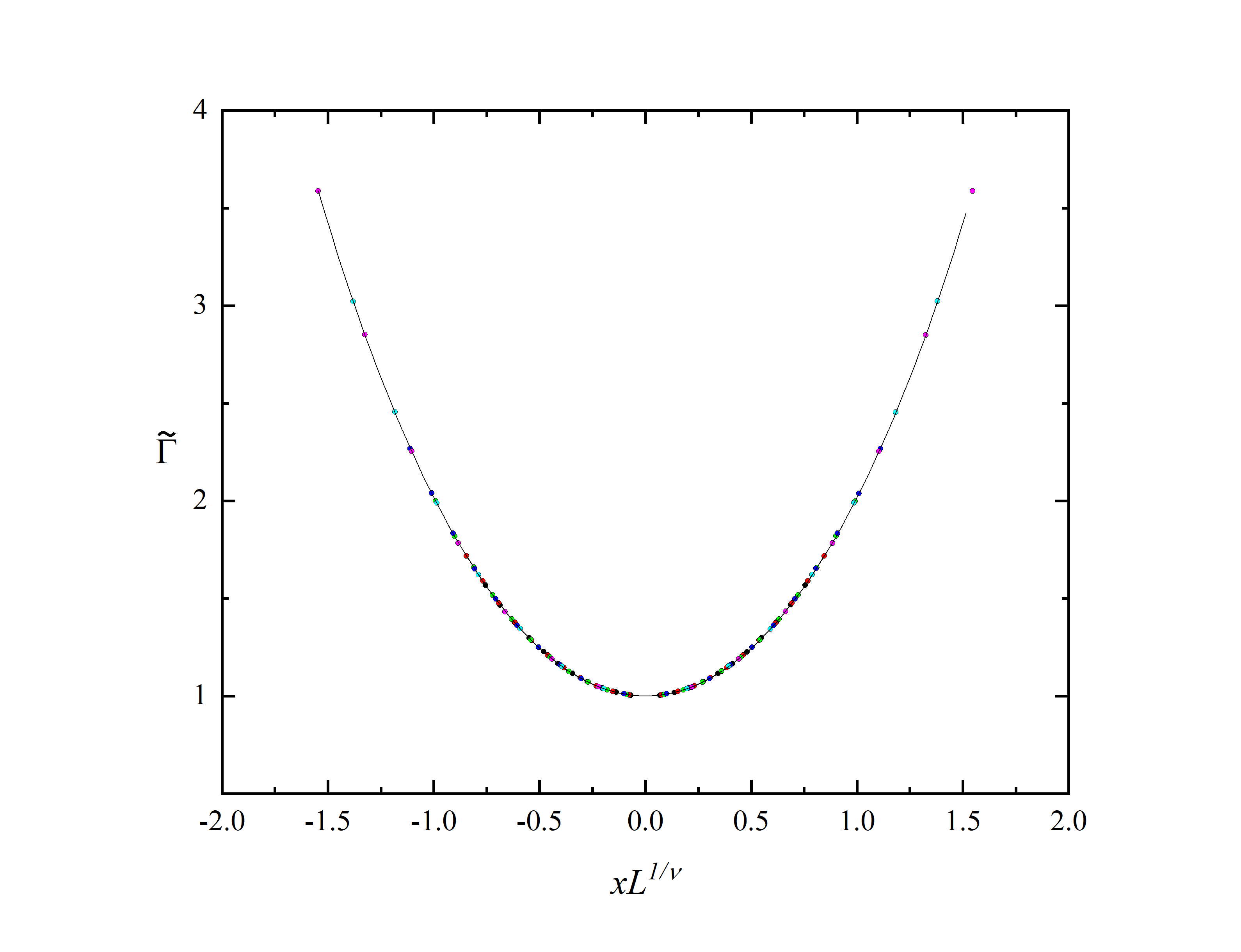}
 \end{center}
  \caption{Data (points) and scaling function (line) to demonstrate single parameter scaling for the Gaussian process fitting.}
\end{figure}

\section{Discussion}

Dresselhaus et al. proposed a factorisation ansatz for the scaling function used in finite size scaling analysis 
of the quantum Hall transition.
We applied this method to our previously published data for the smallest Lyapunov exponent of the
Chalker-Coddington model.
Provided data for smaller system sizes $L$ is excluded, we obtain acceptable fits of our data.
The standard deviations quoted in Table 1 and Table 2 are statistical errors, i.e., they assume there are no systematic deviations between
the model and data.
Looking at Table 1 we see that the estimates of $\nu$ which include non-linearity of $\phi$ agree within the limits of error, i.e., intervals with $\pm 2\sigma$ overlap.
However, the estimates where non-linearity are neglected are of the order of $1\%$ lower.
All the estimates we obtained here are lower than the estimates in the range $2.59$ to $2.61$ obtained 
in our previous work Refs.~\cite{Slevin09,Slevin12} with the same data
where the factorisation ansatz of Dresselhaus et al. was not used.
While these discrepancies are small, we think this indicates that the factorisation ansatz while giving acceptable fits for
some ranges of data is possibly not exact.

\medskip

\medskip
\textbf{Acknowledgements} \par 
K. S. and T. O. were supported by JSPS KAKENHI Grants No. 19H00658,
and T. O. was supported by JSPS KAKENHI Grants No. 22H05114.
K. S and T. O. would like to thank E. Dresselhaus, I. Gruzberg
and Kenji Harada for fruitful discussion.

\appendix

\section{Error calculation}
For our Lyapunov simulation the data are independently normally distributed random variables, i.e., the output of a 
simulation is of the form
\begin{equation}
    \Gamma \left(x, L\right) = \mu\left(x, L\right) + \sigma \eta \,,
\end{equation}
where $\eta$ is a normally distributed random variable with mean zero and variance unity.
The output of the simulation is the estimate $\Gamma$ of the expectation $\mu$ and an estimate of the
standard deviation $\sigma$ of that estimate.
To perform the scaling analysis proposed by Dresselhaus et al., we need to estimate
\begin{equation}
    \frac{\mu\left(x, L\right) }{\mu\left(y, L\right) } \,,
\end{equation}
using data from two independent simulations performed for the same system size $L$ with different energies $x$ and $y$.
The obvious estimate is
\begin{equation}
    \frac{\Gamma\left(x, L\right)}{\Gamma\left(y, L\right)} = \frac{\mu_x+\sigma_x \eta_x}{\mu_y+\sigma_y \eta_y} \,.
\end{equation}
where we abbreviate $\mu\left(x, L\right)$ as $\mu_x$ etc. For all our simulations
\begin{equation}
    \sigma \ll \mu \,.
\end{equation}
Therefore, it is reasonable to expand the left hand side as follows
\begin{equation}
    \frac{\Gamma\left(x, L\right)}{\Gamma\left(y, L\right)} \approx 
    \left(\mu_x+\sigma_x \eta_x\right) \frac{1}{\mu_y} \left(1 + \left( \frac{\sigma_y \eta_y}{\mu_y} \right)^2 + \cdots\right) \,.
\end{equation}
Taking the expectation value we find
\begin{equation}
    \left< \frac{\Gamma\left(x, L\right)}{\Gamma\left(y, L\right)} \right> \approx 
    \frac{\mu_x}{\mu_y} \left(1 + \left( \frac{\sigma_y }{\mu_y} \right)^2 + \cdots\right) \,.
\end{equation}
We see that the bias in the estimate can be neglected for our data.
To fit the data we also need an estimate for the standard deviation of this estimate. 
For the variance we find
\begin{equation}
    {\rm Var} \left( \frac{\Gamma\left(x, L\right)}{\Gamma\left(y, L\right)} \right) \approx
    \frac{\mu_x^2}{\mu_y^2} \left( \frac {\sigma_x^2}{\mu_x^2} + \frac{\sigma_y^2}{\mu_y^2} \right) \,,
\end{equation}
to the same order of approximation.
Thus the standard deviation we need is
\begin{equation}
    \sigma_{x/y}= \frac{\Gamma\left(x, L\right)}{\Gamma\left(y, L\right)} \sqrt{\frac {\sigma_x^2}{\mu_x^2} + \frac{\sigma_y^2}{\mu_y^2}} \,.
\end{equation}

\medskip

%
\bibliographystyle{MSP}
\bibliography{refPSS}


\end{document}